\documentclass[aps,twocolumn,superscriptaddress,showpacs,nofootinbib]{revtex4}
\usepackage{graphicx}
\usepackage{dcolumn}
\usepackage{epsfig}
\usepackage{amsmath}

\begin{document}

\title{Relative stability of $6H$-SiC$\{0001\}$ surface terminations
and formation of graphene overlayers by Si evaporation}

\author{Jochen Rohrer}
\email{rohrer@chalmers.se}
\affiliation{%
BioNano Systems Laboratory,
Department of Microtechnology,
MC2,
Chalmers University of Technology,
SE-412 96 Gothenburg
}%
\author{Eleni Ziambaras}%
\affiliation{%
Materials and Surface Theory Group,
Department of Applied Physics,
Chalmers University of Technology,
SE-412 96 Gothenburg
}%

\author{Per Hyldgaard}%
\affiliation{%
BioNano Systems Laboratory,
Department of Microtechnology,
MC2,
Chalmers University of Technology,
SE-412 96 Gothenburg
}%

\date{\today}

\begin{abstract}
We present density functional theory (DFT) calculations for $6H$-SiC$\{0001\}$ surfaces
with different surface stackings and terminations.
We compare the relative stability of different $(0001)$ and $(000\bar1)$
surfaces in terms of their surface free energies.
Removing surface and subsurface Si atoms, we simulate the formation of
graphene and graphene-like overlayers by Si evaporation.
We find that overlayers with a different nature of bonding are
preferred at the two non-equivalent surface orientations.
At $(0001)$, a chemically bonded, highly strained and buckled film is predicted.
At $(000\bar1)$, a van der Waals (vdW) bonded overlayer is preferred.
We quantify the vdW binding and show that it can have a doping effect on 
electron behavior in the overlayer.
\end{abstract}


\maketitle

\section{Introduction}
Graphene \cite{ref:Graphene} is a highly interesting material
for future nanoelectronic devices \cite{Rise1,Rise2}
such as transistors \cite{Trans},
integrated circuits \cite{IntegratedCircuit}
or  detectors \cite{Detect}.
This is because of its unique electronic properties but also,
and more importantly,
because of the possibility to  adjust and control these properties.
The nature and magnitude of conduction
and overall (opto-) electronic properties
can be modified,
for example,
by applying electric fields \cite{ElField},
by adsorbants \cite{ChemMod1,ChemMod2},
by utilizing finite-size effects \cite{NanoRibbons1, NanoRibbons2}
or by an environment-induced material transformation
into graphane \cite{Sofo, Elias}
or graphene oxide \cite{C2O2}.

An important issue for the use of  graphene and graphene-derivatives \cite{GrapheneDerivatives}
for devices is the actual influence of  substrates
(on which these materials are placed in the device)
on the bandstructure and hence on electron behavior.
Although graphene  and its derivatives 
do not tend to easily form chemical bonds to other materials,
van der Waals (vdW) interactions will always be present.
For the fully hydrogenated graphene-derivative, 
graphane \cite{Sofo, Elias}, 
we \cite{ref:Rohrer_Graphane} have recently reported first-principles 
vdW density functional  (vdW-DF)
calculations \cite{vdW-DF_review},
predicting that vdW interactions stabilize
multilayer formation.
Moreover, we predict that the electron behavior in the 
graphane multilayer may deviate,
at least locally,  
from the behavior in the monolayer \cite{ref:Rohrer_Graphane}.
Similar effects must naturally also be expected 
when graphene is simply exposed to a substrate.
Indeed, combinations of vdW-DF and GW \cite{Hedin} calculations 
for graphene on various metal surfaces 
have already predicted that the vdW interaction can shift the graphene Fermi level \cite{Jacobsen}
and that these shifts can be either positive or negative, depending on the
the actual substrate material.

In this paper, 
we study graphene and graphene-like overlayers at 6H-SiC$\{0001\}$ surfaces,
focusing on vdW binding and the effect of vdW forces on electron behavior.
The 6H-SiC$\{0001\}$ surfaces are a natural choice 
for studying general substrate-graphene vdW interactions.
SiC  is regarded as promising substrate candidate 
for large-scale fabrication of pure graphene 
by Si evaporation from $\{0001\}$ surfaces
\cite{
PhysRevB78.245403.2008,
JPhysCondensMatter21.134016.2009,
SurfSci603.L87.2009}.
We first characterize the thermodynamic stability of SiC$\{0001\}$
surfaces with different orientations, atomic stacking and surface termination.
We then simulate the Si evaporation be removing Si atoms 
from surface and subsurface layers,
letting the systems find their new ground state.
At $(0001)$, we predict a chemically bonded, strongly buckled and stretched
graphene-like overlayer.
At $(000\bar1)$, we predict a flat vdW-bonded graphene overlayer.
For the vdW-bonded overlayer we perform band-structure calculations
and find a modified electron behavior 
indirectly induced through  vdW forces.

The paper is organized as follows.
In Sec.~\ref{Background}, we give a short background
of SiC and its $\{0001\}$ surfaces.
We present details of our computational method in Sec.~\ref{Method}.
In Sec.\ref{Sec:Results} we present our results
for the stability of various surfaces and surface/overlayer systems,
vdW binding between SiC and graphene
and the resulting band structure.
The results are discussed in Sec.~\ref{Discussion}
and we summarize and conclude our work in Sec.~\ref{Summary}.

\section{Materials Background\label{Background}}

\begin{table}[b]
\begin{ruledtabular}
\begin{tabular}{ccc}
        & exp (Ref.~\onlinecite{SiCLatticeExp})& present\\
$a$ in \AA &  3.073             &3.091\\
$c$ in \AA &  15.118            &15.181
\end{tabular}
\end{ruledtabular}
\caption{\label{tab.Bulk}
Comparison of SiC bulk lattice parameters
from experiment and our first-principle calculations.} 
\end{table}

\subsection{Bulk SiC}

\begin{figure}[t]
\includegraphics[width=7.5cm]{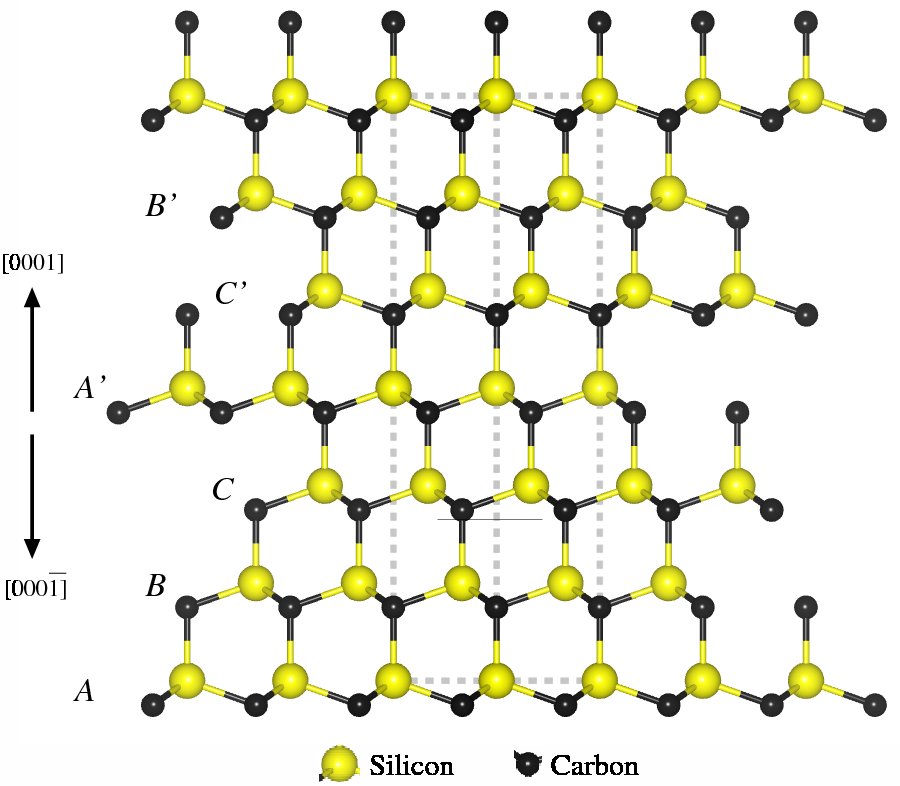}
\caption{
\label{fig:bulk}
(Colour online)
Bulk  structure of 6H-SiC.
Si atoms are represented by yellow (light gray), large spheres;
C atoms are represented by black (black), small spheres.
The light gray dashed lines represent the unit cell.
}
\end{figure}

Figure~\ref{fig:bulk} details the atomic structure of $6H$-SiC. 
The crystal structure is hexagonal. 
Table~\ref{tab.Bulk} presents a comparison of the lattice parameters
obtained from experiment \cite{SiCLatticeExp}
and  from our first-principle calculations
described further below.

Along the $c$-axis, the bulk repeat unit is composed of six SiC bilayers.
Each bilayer contains 50~\% silicon and 50~\% carbon
and forms a buckled honeycomb lattice.
We define the $[0001]$ direction as indicated in the figure.
With this definition, the Si atoms in each bilayer are 
located above (along $[0001]$) the center-of-mass plane,
C atoms are located below.

The stacking sequence (along $[001]$) is $ABCA'C'B'$.
We use primes to distinguish the first $A$, $B$, and $C$ layers 
in each repeat unit from the second ones.
Because of the six layers in the repeat unit,
naively one would expect twelve possible surface terminations
for each surface orientation.
Due to symmetry, however, each two of the twelve terminations per
surface are equivalent.
A rotation by 180$^{\circ}$ around the $c$ axis, 
maps  $A$ sites on $A$ sites,
$B$ sites on $C$ sites and $C$ sites on $B$ sites.
Therefore, the rotated stacking is $A'C'B'ABC$
showing the equivalence of $A$ and $A'$ sites,
$B$ and $C'$ sites, and $C$ and $B'$ sites.

\subsection{Surfaces}
\begin{figure}
\begin{tabular}{ccc}
\includegraphics[width=8.4cm]{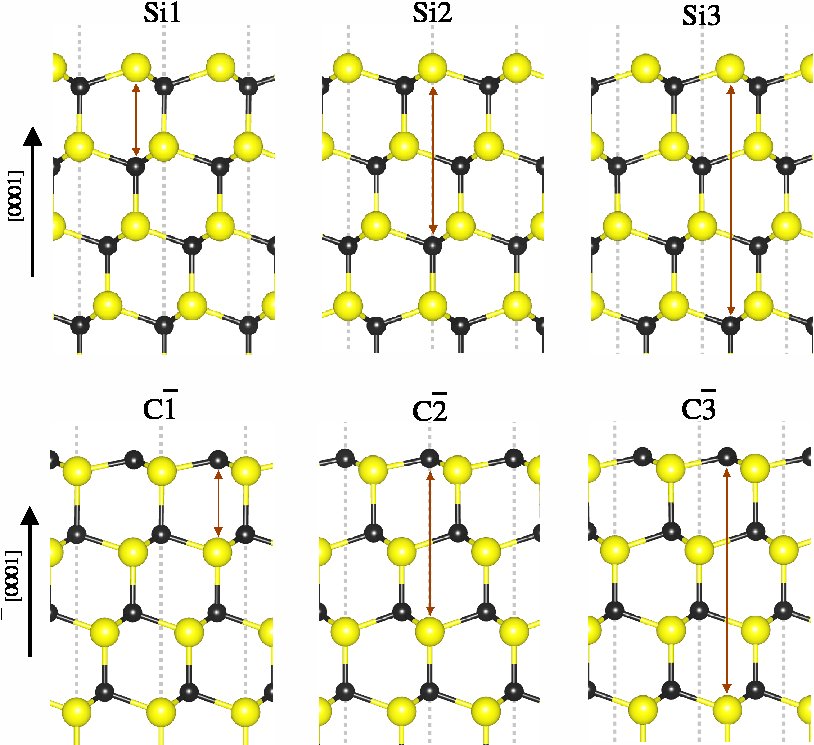}
\end{tabular}
\caption{
\label{fig:surfaces}
(Color online)
$(1\times1)$ 6H-SiC$(0001)$ and $(000\bar{1})$ surfaces
with different details stacking in the outermost layers.
Color coding as in Fig.~\ref{fig:bulk}.
The set of  top panels shows the three nonequivalent
$(0001)$ surfaces with Si termination Si$1$, Si$2$ and Si$3$
(see text for a definition of this labeling).
Corresponding C-terminated surfaces (C$1$, C$2$ and C$3$) are obtained
by adding a C layer on top of terminating Si layer.
The set of bottom panels shows the three nonequivalent
$(000\bar{1})$ surfaces with C termination C$\bar{1}$, C$\bar2$ and C$\bar3$.
Corresponding Si-terminated surfaces 
(Si$\bar1$, Si$\bar2$ and Si$\bar3$) are obtained
by adding a Si layer on top of the terminating C layer.
}
\end{figure}

Figure~\ref{fig:surfaces} shows six out of the twelve 
different (ideal) $6H$-SiC$\{0001\}$ surface configurations.
Along the $[0001]$ direction, the $6H$ crystal structure lacks inversion symmetry.
Thus, the corresponding $(0001)$ surface (the so-called nominally Si-terminated surface \cite{Nominally})
and $(000\bar{1})$ surface (the so-called nominally C-terminated surface \cite{Nominally}) are different.
For each surface orientation there are six different possible ideal terminations,
corresponding to two  chemically different terminations (Si or C)
times three structurally different terminations.
Ideal here means that we only consider full-coverage surfaces.
In practice, a large surface may exhibit partial coverage
to counteract a diverging surface dipole \cite{Tasker}
and there may be surface reconstructions as to 
saturate surface dangling bonds.

The top panels show the set of three most natural (flat)
Si-terminated   $(0001)$ surfaces. 
These are denoted by Si$1$, Si$2$, and Si$3$.
In a Si$1$ surface, the surface Si atoms are located on top of 
C atoms that are located in the first subsurface SiC bilayer.
In Si$2$ and Si$3$ surfaces, the surface Si atoms are correspondingly
located on top of C atoms in the second and third subsurface SiC bilayer.
The C-terminated counterparts (C$1$, C$2$, and C$3$)
are obtained by adding an additional C layer on top of the surface Si layer.

The set of bottom panels displays C-terminated $(000\bar1)$ surfaces.
The three surfaces are denoted by C$\bar1$, C$\bar2$ and C$\bar3$.
In a C$\bar1$ surface, the surface C  atoms are located on top of 
Si atoms that are located in the first subsurface SiC bilayer.
In C$\bar2$ and C$\bar3$ surfaces, the surface C atoms are correspondingly
located on top of Si atoms in the second and third subsurface SiC bilayer.
The Si-terminated counterparts (Si$\bar1$, Si$\bar2$, and Si$\bar3$)
are obtained by adding an additional
Si layer on top of the surface C layer.

At the $(0001)$ surface, C atoms bind only to a single Si atom below;
they possess three dangling bonds.
Si atoms, on the other hand, bind to three nearest-neighbor C atoms;
they only possess one dangling bond.
At $(000\bar1)$, the situation is reversed.
Therefore, in the absence of complex reconstructions,
the  $(0001)$ surface is intuitively expected  to be Si terminated
(and therefore denoted as the nominally Si-terminated surface \cite{Nominally})
and the $(000\bar1)$ surface is intuitively expected to be C terminated
(and therefore denoted as the nominally C-terminated surface \cite{Nominally}).

\section{Computational Method\label{Method}}

\subsection{Surface and overlayer stability}
All surface and surface/overlayer
calculations are performed with the planewave pseudopotential \cite{Vanderbilt} 
DFT code \textsf{dacapo} \cite{Dacapo}
and the PBE \cite{PBE} functional for exchange and correlation.
We use a planewave cutoff of $400$~eV
and a $(4\times4\times1)$ k-point sampling \cite{MonkhorstPack} 
Force relaxations are performed until the residual force
is less than $0.03$~eV/\AA.

We use the supercell approach and represent each surface or
surface/overlayer system by slab geometry.
The supercells have a height of $40$~\AA\ and
the lateral dimensions are fixed 
to accommodate the $(1\times1)$ SiC$\{0001\}$ surface.
Surfaces with different stackings and chemical terminations 
are represented by slabs of different thicknesses,
varying from 9 and 15 bilayers (plus one optional excess Si or C layer).
Also, we saturate dangling bonds at 
the $(000\bar1)$  [$(0001)$] side of the slab that represents
the $(0001)$ [$(000\bar1)$] surface
by attaching H atoms.
Because of the asymmetry of the slabs, 
we use a dipole correction \cite{Bengtsson}.
With respect to thickness, 
the calculated surface-energy differences, see Table~\ref{tab:DeltaSigma}, 
are converged by at least $\pm2$~meV/\AA$^2$.
We have  tested this accuracy by comparing surface energies
of equivalent surfaces that are represented by slabs of different thicknesses.
Specifically, any surface represented by a particular slab
can be recovered by addition of three SiC bilayers.

\subsection{Thermodynamic comparison of chemically different terminations}
We determine the equilibrium surface preference 
by comparing surface free energies \cite{AIT-SE1, AIT-SE2}.
Due to the lack of inversion symmetry along $[0001]$,
use of slab geometry only enables us to calculate the sum
(denoted by $\tilde{\sigma}$ and defined below) of the two surface energies (denoted by $\sigma$)
corresponding to the   $(0001)$ surface and the $(000\bar1)$ surface.
However, keeping the geometry and chemical composition
fixed at one side of the slab,
we ensure that the contribution to $\tilde{\sigma}$ from that side is always the same.
We thus can compare the relative stability of different structures and compositions
at the other side (not fixed) by considering the difference in  $\tilde{\sigma}$
for various stackings and terminations.

We define the sum of the two surface free energies as
\begin{align}
\tilde{\sigma} = 
\frac{1}{A}\bigg(E_{\text{slab}}-n_{\text{Si}}\epsilon_{\text{SiC}}
-(n_{\text{C}}-n_{\text{Si}})\mu_{\text{C}}-n_{\text{Gr}}\epsilon_{\text{Gr}}\bigg).
\label{eq:sigma}
\end{align}
Here, $E_{\text{slab}}$ is the total energy of the $6H$-SiC 
surface slab, including a possible graphene-like overlayer,
that contains $n_{\text{Si}}$ silicon atoms (per supercell), 
$n_{\text{C}}$ carbon atoms that belong to the SiC
and $n_{\text{Gr}}$  carbon atoms that belong to the graphene overlayer.
Furthermore, $\epsilon_{\text{SiC}}$ denotes the energy of one stoichiometric unit of bulk SiC,
$\epsilon_{\text{Gr}}$ is the energy of graphene,\footnote{
This energy must contain the strain energy  since
the graphene overlayer is highly expanded in a SiC surface cell.}
and $\mu_{\text{C}}$ is carbon chemical potential.
We note that we have assumed equilibrium between the bulk and the surface,
$\epsilon_{\text{SiC}}=\mu_{\text{Si}}+\mu_{\text{C}}$.

The values of the carbon chemical potential in (\ref{eq:sigma})
are restricted to a finite range.
If the carbon chemical potential is larger than the free energy per carbon atom
in graphene ($\mu_{\text{C}}>g_{\text{graphene}}$), 
the formation of (additional) graphene overlayers becomes more favorable.
If, on the other hand,
the silicon chemical potential is larger than the free energy per silicon atom
in bulk Si (with diamond structure, $\mu_{Si}>g_{\text{Si}}$),
the formation of bulk Si is more favorable.
Therefore, the allowed range of the carbon chemical potential is
\begin{align}
g_{\text{graphene}}\geq\mu_{C} \geq g_{\text{SiC}}-g_{\text{Si}}.
\end{align}

We also introduce
\begin{align}
\Delta \sigma_{ij} =\tilde{\sigma}_i-\tilde{\sigma}_j,
\end{align}
from which we infer the relative stability of surfaces with identical orientations
but different stackings and chemical compositions.

\subsection{van der Waals binding}
In Sec.~\ref{Overlayers},
we identify a chemically non-binding SiC$(000\bar1)$/graphene
system as thermodynamically stable.
We study this system further using the van der Waals
density functional (vdW-DF) method \cite{vdW-DF_review}.
In particular we use the nonlocal correlation functionals 
$E^{\text{nl-}1}_{\text{c}}$ and $E^{\text{nl-}2}_{\text{c}}$
of  Refs.~\onlinecite{vdW-DF1} and \onlinecite{vdW-DF2}, respectively.

We calculate the energy variation, including vdW forces,
by a postprocessing method as follows.
We first perform traditional DFT calculations (using the
PBE exchange-correlation functional)
for various separations between the SiC and the graphene overlayer.
In these calculations we choose the lateral dimensions 
of the unit cell to fit those of a $(4\times4)$ SiC$(000\bar1)$ surface.
This allows us to study a $(4\times4)$ SiC$(000\bar1)$/$(5\times5)$ graphene
system in which the graphene overlayer is hardly strained at all.
We use a $(3\times3\times1)$ k-point sampling
to ensure an accurate electronic density for further  evaluations
of nonlocal correlations according to 
\begin{align}
E^{\text{vdW-DF-}v}[n]=E_{0}^v[n]+E^{\text{nl-}v}_{\text{c}}[n].
\label{eq:vdWDF}
\end{align}
Here, $E^{\text{nl-}v}_{\text{c}}[n]$ is the correlation energy from one of 
the non-local functionals  of Refs.~\onlinecite{vdW-DF1} ($v=1$) and \onlinecite{vdW-DF2} ($v=2$).
Also, in Eq.~(\ref{eq:vdWDF}), $E_{0\text{-}v}[n]$ is given by
\begin{align}
E_{0}^v=E^{\text{PBE}}_{\text{tot}}-E^{\text{PBE}}_{\text{xc}}+E^{\text{VWN}}_{\text{c}}+
E^{v}_{\text{x}},
\label{eq:Enl}
\end{align}
where $E^{\text{VWN}}_{\text{c}}$ is the VWN-LDA \cite{VWN} correlation energy
and $E^{v=1}_{\text{x}}= E^{\text{revPBE}}_{\text{x}}$
and $E^{v=2}_{\text{x}}= E^{\text{rPW86}}_{\text{x}}$
are the revPBE  \cite{revPBE} and rPW86 (refitted form of PW86) \cite{rPW86})
exchange functionals.

We determine the vdW binding separation and energy
by finding the minimum in the layer-binding energy 
defined as
\begin{align}
E_{\text{bind}}(d)&=E_{\text{vdW-DF}}(d)-
E_{\text{vdW-DF}}(d\rightarrow\infty).
\label{eq:Ebind}
\end{align}
Here, $d$ is the distance between the 
surface layer and the overlayer.
For a detailed description of a robust implementation
of the evaluation 
of Eq.~(\ref{eq:Ebind}), 
we refer to Refs.~\onlinecite{sliding,Method1,Method2,Mehtod3}.

\subsection{Band-structure calculations}
We also perform band-structure calculations for  $(5\times5)$ graphene on $(4\times4)$ SiC$(000\bar1)$
to probe the effect of vdW bonding on electron behavior.
The band-structure calculations are performed as follows.
We fix the  SiC-graphene separation 
at the value predicted by the vdW-DF2 calculations
and determine the density with a $(4\times4\times1)$ k-point sampling
to ensure a high accuracy in our large unit-cell calculations.
This density is subsequently used to 
perform traditional GGA calculations of the energy spectra 
at various k points.
We focus on the  special Brillouin-zone points  
$\Gamma=(0,0,0)$, $\text{K}=(2/3, 1/3,0)$
and $\text{M}=(1/2,1/2,0)$
and the lines
along $\overline{\text{K}\Gamma}$, $\overline{\Gamma\text{M}}$,
and $\overline{\text{K}\text{M}}$.

Since we are interested in band-structure modifications,
we also perform the same type of calculations for a single graphene layer.
In order to not encounter modifications that may be solely due
to a (very small) variation in the graphene lattice parameter,
we use the same unit cell as in the case for the SiC/graphene system.

\section{Results \label{Sec:Results}}

\subsection{Relative stability of clean $6H$-SiC$\{0001\}$ surfaces\label{Surfaces}}
\setcounter{paragraph}{0}

\begin{table}[t]
\begin{ruledtabular}
\begin{tabular}{lrlr}
$(0001)$     &$\Delta\sigma$ [meV/\AA$^2$]&  $(000\bar1)$  &$\Delta\sigma$ [meV/\AA$^2$]\\[0.2cm]
Si1&     6.7  &    Si$\bar1$&  1.9  \\
Si2&     0    &    Si$\bar2$&  0  \\
Si3&     1.7  &    Si$\bar3$&  0.3 \\
C1&      4.6  &    C$\bar1$&   0  \\
C2&      0    &    C$\bar2$&   3.0  \\
C3&      1.3  &    C$\bar3$&   3.5 
\end{tabular}
\end{ruledtabular}
\caption{
\label{tab:DeltaSigma}
Differences in surface energies for surfaces with identical orientations
and terminations (which makes the numbers independent of the chemical potentials)
but with different detailed stacking of the outermost surface layers.
The energetically most favorable surface for each orientation and termination 
is chosen as reference and therefore characterized  by $\Delta\sigma = 0$.
The effect of surface vibrations is not included.
}
\end{table}

\paragraph{Stacking preference for fixed chemical termination. }
Table~\ref{tab:DeltaSigma} compares surface energies for 
surfaces with identical orientations and identical chemical compositions
but with a different detailed stacking of the surface layers.
For each orientation and chemical composition of the surface,
we choose the configuration with lowest surface energy,
or more precisely with lowest $\tilde{\sigma}$,
as reference.
This configuration is thus characterized by  $\Delta\sigma = 0$.

We find that the surface-energy differences are of the order of a few meV/\AA$^{2}$.
These values are close to or below the actual accuracy of our calculations.
Thus, the surface-energy differences between 
surfaces with different stackings
are too small to be of significant importance for
determining the stable surface configuration.

\begin{figure*}
\begin{center}
\includegraphics[width=17.5cm]{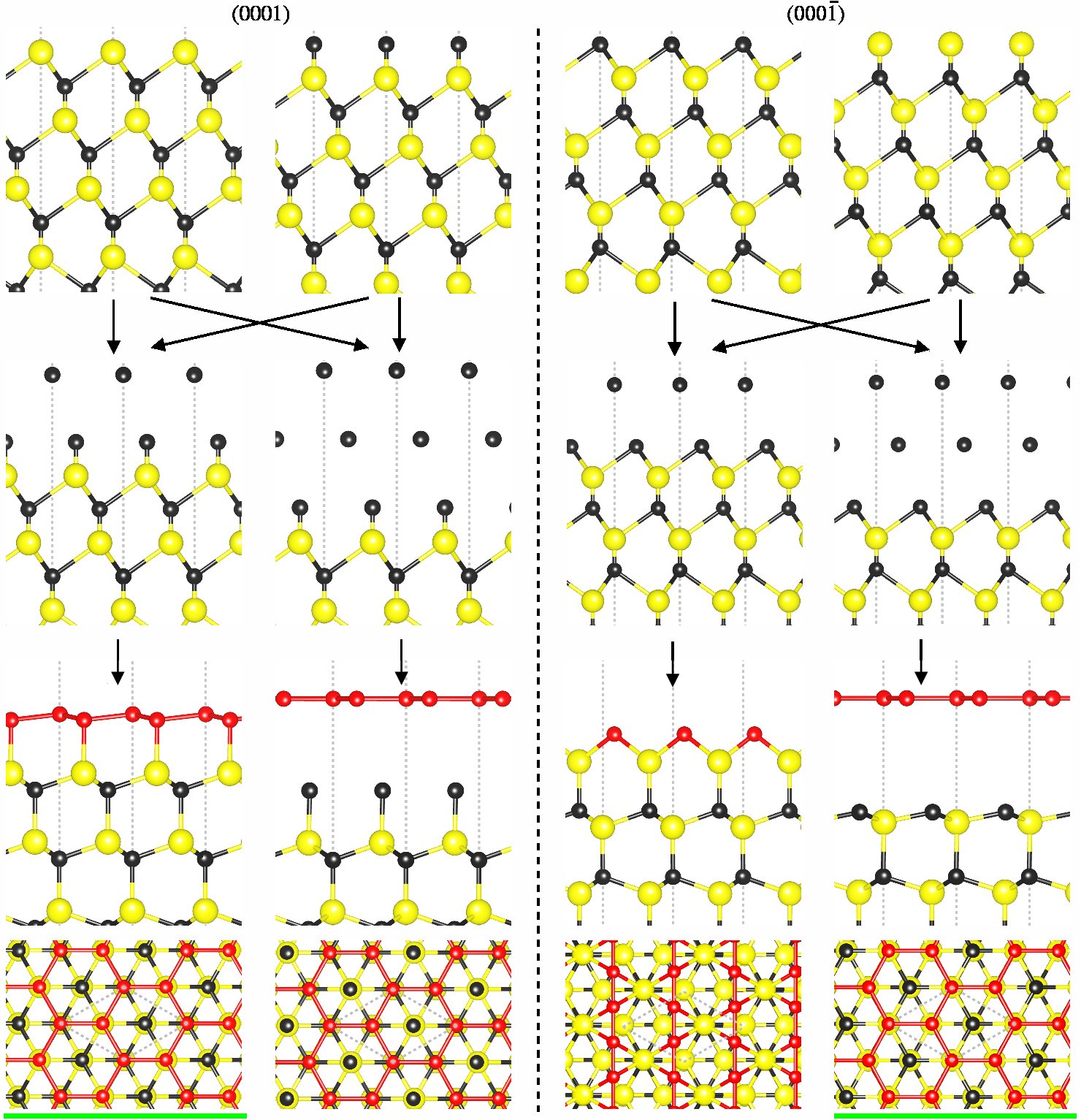}
\end{center}
\caption{
\label{fig:Vacancies}
(Color online)
Formation of graphitic overlayers at SiC$\{0001\}$.
Color coding as in Fig.~\ref{fig:bulk};
in addition, C atoms forming an overlayer are in red (dark gray)
and connected through bonds.
The set of top panels shows side views on truncated $(0001)$ and $(000\bar1)$ surface structures with either Si or C termination.
The set of mid-panels shows side views the same structures (\textit{i.e.} unrelaxed),
but with (sub) surface Si layers removed.
The set of bottom panels shows side and top views on the  geometries obtained by
relaxing the systems in the mid-panels.
At each surface there exist chemically bonded systems (first and third column in the bottom set of panels) 
and systems that are not chemically bonded (second and fourth set of panels).
Our thermodynamics analysis indicates that the
chemically bonded system is the stable one at SiC$(0001)$;
At  SiC$(000\bar1)$, the chemically non-binding configuration is stabilized.
The stable, that is, thermodynamically favored  
configurations are underlined  in green (gray). 
}
\end{figure*}

\begin{table}
\begin{ruledtabular}
\begin{tabular}{lrrr}
                                 & \multicolumn{3}{c}{$\sigma_{\text{Si}} - \sigma_{\text{C}}$ in meV/\AA$^2$ at}\\
                                 &$\mu_{\text{C}}^{\text{min}}$&$\mu_{\text{C}}^{\text{mid}}$ & $\mu_{\text{C}}^{\text{max}}$\\[0.2cm]
1$\times$1 $(0001)$              &   -527&  -502 & -476 \\
$\sqrt{3}\times\sqrt{3}$ $(0001)$&   -559&  -534 & -509 \\
3$\times$3 $(0001)$              &   -539&  -513 & -487 \\[0.1cm]
1$\times$1 $(000\bar1)$          &   -128&  -102 &  -77  \\
$\sqrt{3}\times\sqrt{3}$ $(000\bar1)$  &    -75&  -49 &  -23 \\
3$\times$3 $(000\bar1)$           &   -118  &  -93  &  -67\\
\end{tabular}
\end{ruledtabular}
\caption{
\label{tab:Sigma}
Preference of surface chemical composition at  $6H$-SiC$\{0001\}$.
The table lists differences in surface free energies between Si- and C-terminated surfaces
for three values of the chemical potential of C.
A negative value indicates higher stability of the 
Si-terminated surface.
}
\end{table}

\paragraph{Chemical composition. \label{sec:ChemicalComposition}}
In Table \ref{tab:Sigma} we determine the preferred chemical composition
of  the $(0001)$ and $(000\bar1)$ surfaces.
For each orientation
we list differences in surface free energies between Si- and C-terminated surfaces
at three different values of the C chemical potential.
The three values of the C chemical potential correspond to
its maximal value, its minimal value, and the value in between.

We find that, within the entire allowed range of the C chemical potential, 
the Si-terminated surfaces possesses lowest free surface energy,
independent of the orientation.
For $(0001)$ orientation, this coincides with the expectations
on the basis of the number of dangling bonds.
For $(000\bar1)$ surfaces, the predictions appear counterintuitive.
There, the Si-terminated surface possesses more dangling bonds
than the C-terminated surface
and should therefore be less favorable.

Table~\ref{tab:Sigma} also reports that
the perhaps counterintuitive result for $(000\bar1)$ surfaces
remains unaltered when considering larger surface unit-cells
that allow for surface reconstructions (without considering
more complex surface terminations than pure Si or C termination, however), 
such as $(\sqrt{3}\times\sqrt{3})$ and $(3\times3)$ surface unit-cells.
Details of our calculations concerning
surface reconstructions are documented in the
supplementary material.

The appearant equilibrium preference of Si-terminated 
$(000\bar1)$ surfaces
reflects the rich phase diagram of SiC.
Our comparison, limited to full Si and C coverages,
is likely not exhaustive enough to capture this richness
in a more quantitative manner.
On the other hand, 
resolving the difficult structure is not 
relevant for our search for carbon overlayers
at SiC$\{0001\}$ surfaces
that may form by evaporation of Si atoms.

\subsection{Graphitic overlayer formation by Si evaporation\label{Overlayers}}
\setcounter{paragraph}{0}
We now turn our focus towards structure, stability and bonding of
graphitic overlayers at SiC$\{0001\}$.
Figure  \ref{fig:Vacancies} illustrates the formation of different 
graphitic overlayers at SiC$\{0001\}$ surfaces
by evaporation of Si atoms \cite{PhysRevB78.245403.2008,JPhysCondensMatter21.134016.2009,SurfSci603.L87.2009}.
We use $(1\times1)$ surface unit-cells to 
study the effect of removing full surface and subsurface layers.
As illustrated in the preceding subsection and the supplementary material, 
the initial surface morphologies can,
in principle,  be much more complicated.
However, since the surface atoms evaporate in a heating process,
we expect that the detailed surface structure before heating 
is of minor importance.

\paragraph{Structure. }
The set of top panels of Fig.~\ref{fig:Vacancies} shows 
four $(1\times1)$ surfaces with different orientations and terminations.
In the set of mid-panels, we have removed surface and subsurface Si atoms.
The set of bottom panels presents the geometries that are obtained 
by relaxing the structures in the set of mid-panels above.
The figure also illustrates that the role of the surface composition (and structure)
effectively reduces to determining the amount of Si that needs
to be evaporated to generate a specific final SiC/graphene system.

At both surface orientations we obtain two types of overlayers.
The first type of overlayer is chemically bonded,
see first and third column 
in the set of bottom panels in Fig.~\ref{fig:Vacancies}.
As a result the overlayer adopts the SiC lattice parameter
resulting in a considerable stretching of the C bonds.
At $(0001)$, the overlayer preserves the hexagonal graphene-like shape
but also exhibits a buckling.
At $(000\bar1)$, the hexagonal shape is not preserved.
The C atoms arrange themselves in chains
located in bridge positions at the surface.

The second type of overlayer is not chemically bonded,
see second and fourth column in the set of bottom panels of Fig.~\ref{fig:Vacancies}.
The non-binding character is reflected  by the large separations between 
the overlayers and the outermost surface layers.
In both cases, the absence of chemical bonding
leaves the ideal hexagonal graphene shape unchanged.
We expect that these carbon overlayers 
contract and adopt the unstrained graphene lattice.

\paragraph{Stability. }
We compare the stability of the different surface/overlayer systems
with identical orientations by means of their surface energy,
see Eq.~(\ref{eq:sigma}).
For the chemically bonded systems, 
we consider the overlayer to be in equilibrium with the SiC surface,
that is, the number of graphene units is set to zero.
For the other two systems, we assume that the chemical potential
of the overlayer is not related to that of SiC,
but to that of (strained) graphene (so correcting for the strain energy
due to the lattice misfit).

At the $(0001)$ surface, we find that
the chemically bonded system is  more preferred.
The surface-energy differences are  86~meV/\AA$^2$ 
at the minimal value of carbon chemical potential
and 137~meV/\AA$^2$ at the  maximal value of carbon chemical potential.
At the $(000\bar1)$ surface, the chemically non-bonding system is preferred.
The corresponding surface-energy differences are 445~meV/\AA$^2$ (at $\mu_{\text{C}}^{\min}$)
and 393~meV/\AA$^2$ (at $\mu_{\text{C}}^{\min}$).

\begin{figure}
\includegraphics[width=8cm]{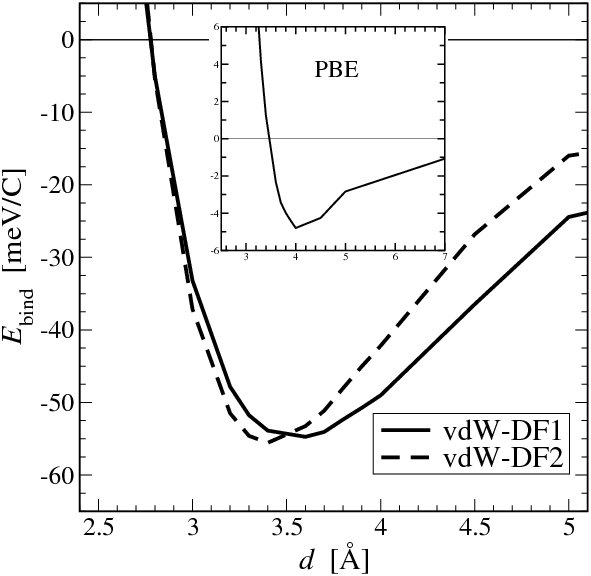}
\caption{
\label{fig:vdW}
Energy variation of graphene on SiC$(000\bar1$) including vdW forces.}
\end{figure}

\begin{figure}[t]
\includegraphics[width=8cm]{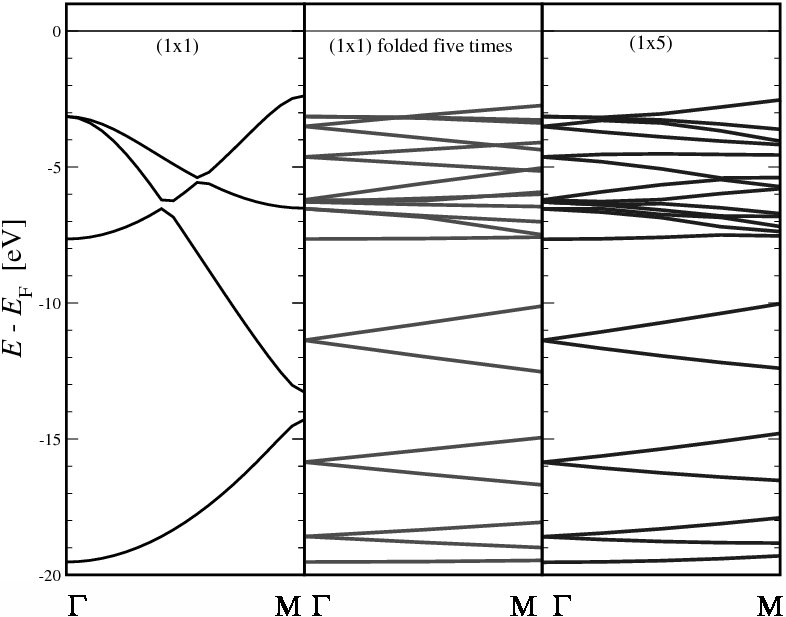}
\caption{
\label{fig:zoneFolding}
Zone folding of the graphene band diagram along $\overline{\Gamma\text{M}}$.
The left panel shows the unfolded calculated $(1\times1)$ band structure of graphene.
The diagram in the mid-panel is constructed from that in the left
by folding five times.
The right panel shows the calculated band diagram using a $(1\times5)$ 
graphene unit cell.
}
\end{figure}

\begin{figure}[b]
\begin{tabular}{c}
\includegraphics[width=8cm]{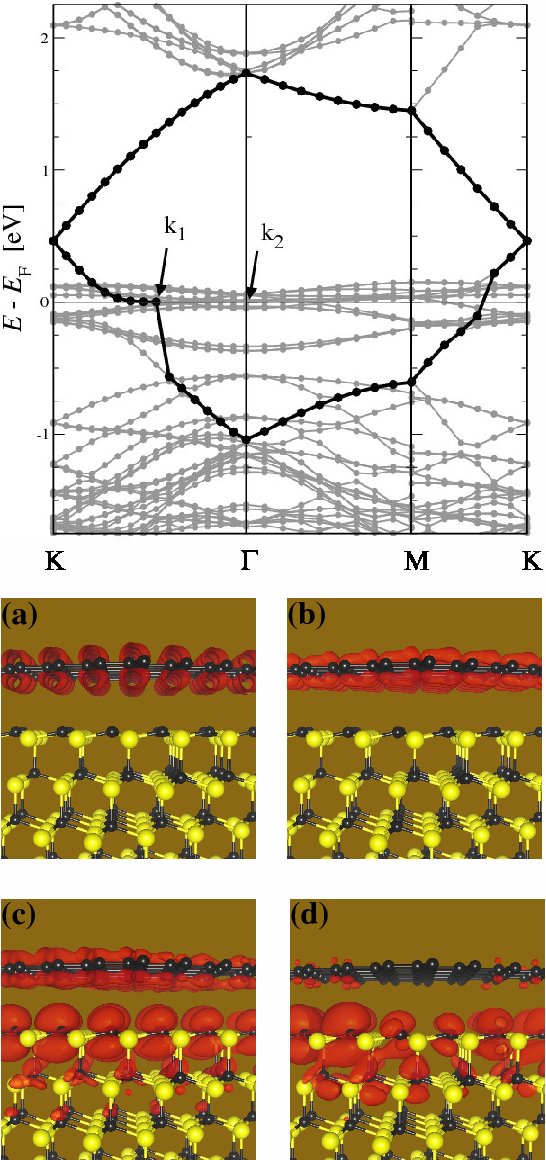}
\end{tabular}
\caption{
\label{fig:BS}
(Color online) Electronic structure of  vdW bonded 
$(5\times5)$ graphene at $(4\times4)$SiC$(000\bar1)$.
The top panel shows the calculated  band diagram
along $\overline{\text{K}\Gamma}$,
$\overline{\Gamma\text{M}}$
and $\overline{\text{M}\text{K}}$.
The UVB and LCB corresponding to graphene are highlighted.
The set of bottom panels shows  (absolute values of) 
various wave functions.
(a) Graphene UVB WF in K, (b) graphene LCB WF in K,
(c) WF in $k_1$ (see top panel)
and
(d) WF in $k_2$ (see top panel).
}
\end{figure}

\subsection{vdW bonded $(5\times5)$ graphene at $(4\times4)$ SiC$(000\bar1)$}
\setcounter{paragraph}{0}
We quantify the vdW binding of graphene at SiC$(000\bar1)$
by studying the vdW energy variation of the system shown in the rightmost
bottom panel of Fig.~\ref{fig:Vacancies}.
In the depicted $(1\times1)$ system, 
the graphene overlayer is highly strained.
We therefore use a  
$(4\times4)$ SiC$(000\bar1)$ surface unit-cell in our calculations
combined with a $(5\times5)$ graphene overlayer \cite{EleniThesis}
to ensure that the overlayer is almost unstrained 
(the C-C bond length is stretched by $\sim0.5$\% only).

Figure~\ref{fig:vdW} shows the energy variation as a function of 
the separation between the SiC surface-layer and the graphene overlayer.
Both, results using vdW-DF1 and vdW-DF2 are shown.
The insert shows that traditional GGA calculations (PBE)
do not predict any meaningful binding (notice the different scale
on the $y$-axis in the insert).

The vdW-DF energy variations agree qualitatively.
vdW-DF2 predicts a slightly  smaller binding separation
and a slightly larger binding energy.
The numerical values are $d_{\text{bind}}=3.6$~\AA\ 
and $E_{\text{bind}}=-54.7$~meV per carbon atom (in graphene)
for vdW-DF1
and $d_{\text{bind}}=3.4$~\AA\ 
and $E_{\text{bind}}=-55.6$~meV per carbon atom 
for vdW-DF2.

We note that accounting for vdW binding 
for the chemically non-binding graphene overlayer
at SiC$(0001)$ (second panel from the left in bottommost set of panels
in Fig.~ ~\ref{fig:Vacancies})
is expected to lower surface energy by a similar amount.
However, our calculated values correspond to $\Delta_{\text{vdW}}\sigma\sim13$~meV/\AA.
Therefore, even with an account of vdW interactions,
the chemisorbed carbon overlayer will still be more preferable 
than the chemically nonbinding (vdW-bonded) overlayer.

\subsection{Band structure of vdW-bonded graphene at nominally C-terminated SiC$(000\bar1)$}
\setcounter{paragraph}{0}

\paragraph{Consistency check for zone folding. }
In Fig.~\ref{fig:zoneFolding} we check that our large unit-cell calculations 
capture and reliably reproduce the details of the electron behavior,
that is, the band-structure physics.
The left panel shows the band structure of graphene (without substrate) 
along a straight line from $\Gamma$ to K
as calculated within a $(1\times1)$ unit cell.
In the mid-panel we show a band diagram that is constructed
from the left panel by zone folding it five times.
The right panel shows the band diagram as calculated 
within a $(1\times5)$ unit cell.

The constructed and the calculated zone-folded band diagram agree
reasonably well.
The slight differences may be due to differences in the underlying
electronic densities that are used in the respective calculations
and which is transferred from the $(1\times1)$ band diagram to the 
zone-folded diagram.

\paragraph{Overlayer band-structure. }
The top panel of Fig.~\ref{fig:BS} shows the calculated band diagram
of $(5\times5)$ graphene at $(4\times4)$ SiC$(000\bar1)$.
We restrict the plot to relevant energy window around the Fermi level.
The complex band structure is due to the 
$(5\times5)$  zone folding of graphene bands
and the $(4\times4)$ zone folding of SiC bands.

Among the many bands, two bands are highlighted.
These bands correspond to the  upper valence band (UVB)
and to the lower conduction  band (LCB)
in an isolated graphene sheet.
We have explicitly checked that the wave functions (WF)
corresponding to the UVB and LCB are localized on the graphene overlayer.

The set of bottom panels of  Fig.~\ref{fig:BS} shows 
various WFs.
These WFs are representative for the different types
of WF localization in the system.
Panels  (a) and (b) show that 
the UVB and LCB WFs in K, for example, are entirely located on the overlayer.
WFs fully localized on graphene are typical for the LCB.

We note that a kink arises in the band structure variation as the graphene band
crosses the Fermi level at $k_1$ 
(see top panel of Fig.~\ref{fig:BS} for a definition of $k_1$).
There we find that the graphene UVB WF can also be shared between the SiC
and the graphene.

Finally, in panel (d), we illustrate that the graphene overlayer also
slightly affects the nature of the SiC states at the Fermi level in $k_2$
(see top panel of Fig.~\ref{fig:BS} for a definition of $k_2$). 
Although most WFs corresponding to a band between the UVB and LCB
are fully localized within the SiC substrate,
at $k_d$ and other k points,
some WFs do have a small weight also on the graphene overlayer.

\section{Discussion\label{Discussion}}

\subsection{Surface stability}
Our calculations of $(1\times1)$ SiC$\{0001\}$,
including some $(\sqrt{3}\times\sqrt{3})$ and $(3\times3)$ reconstructions
in the supplementary material,
predict a preference for Si termination.
This prediction is reasonable for the $(0001)$
surface but surprising for the $(000\bar1)$ surface
for which we would expect a C termination.

Resolving this discrepancy requires a more careful study
that take into account the actual growth conditions
of SiC \cite{AIT-DG}
and/or improves on the description of surface reconstructions.
The latter task would have to consider a richer set
of surface terminations also including excess and deficiency
Si or C.
However, the size of the $(\sqrt{3}\times\sqrt{3})$ and $(3\times3)$ 
surface unit-cells makes a full reconstruction-search
a large project of its own,
requiring systematic structure-search strategies 
such as considering  a larger pool of candidate geometries with
structural motifs \cite{RohrerAluminaStructure} 
of the reconstructed surfaces obtained  here,
evolutionary-type of iteration \cite{GeneticAlg}
or other global structure-search methods \cite{SimAnneal, BasinHop, MinHop}.

Such a search for surface reconstructions is clearly beyond the scope of the present work,
in particular, since the exact morphology of the stable SiC$\{0001\}$ surfaces
is only of minor importance for the main objective of
this paper: the study of modification of electron behavior
in graphene overlayers due to vdW interactions.
At the same time, we emphasize that our calculated surface energies 
are  upper limits of the true surface energies.

\subsection{Nature of binding in SiC/graphene systems} 
We have identified preferred SiC/graphene systems as they
may result by Si evaporation from various SiC$\{0001\}$ surfaces.
Our results indicate that the nature of binding 
at the nominally Si-terminated $(0001)$ surfaces
is different from the nature of binding
at the nominally C-terminated $(000\bar1)$ surface.

At $(0001)$, we have identified a chemisorbed,
strongly buckled graphene overlayer.
At  $(000\bar1)$, the overlayer is stabilized by vdW forces.
The different nature of binding at the two surfaces
may have consequences for the quality of graphene
that is grown by evaporation of Si from different  SiC$\{0001\}$ faces
(at the nominally C-terminated or at the nominally Si-terminated face).

\subsection{Electron behavior in vdW-bonded graphene}
Our band-structure calculations for SiC/graphene, see Fig.~\ref{fig:BS},
show that vdW binding can cause a doping-like effect.
In free-standing graphene, the density of states (DOS) vanishes
at the Fermi level.\footnote{
We have checked that this remains true also if we slightly stretch the C-C bond length
such that a $(5\times5)$ unit cell is commensurate with
a $(4\times4)$ SiC$\{0001\}$ cell.}
On SiC, the vdW binding renders graphene  a p-doped metal.
The Fermi level is shifted
to lower energies into the original valence band
where the DOS is finite.
The prediction of a Fermi-level shift is similar to the
results reported for graphene on various metal surfaces
in Ref.~\onlinecite{Jacobsen}.

In the present case, vdW binding leads to a further
modification in the band-structure.
At the new Fermi level between K and $\Gamma$, 
we also observe a kink in the band structure.
To the left of $k_1$ (see Fig.~\ref{fig:BS})
the dispersion in the LCB can be fitted to a parabolic
form $E(k)=E_0+a\cdot(k-k_1)^2$,
from which we infer an effective electron mass  
\begin{align}
m_{\text{eff}}=\hbar^2\left[\frac{d^2E}{dk^2}\right]^{-1}.
\end{align}
$m_{\text{eff}}\sim1.02\times10^{-30}$~kg or $m_{\text{eff}}\sim1.1m_{\text{e}}$
where $m_{\text{e}}$ is the (actual) electron mass.


\section{Summary and Conclusions\label{Summary}}
We present DFT calculations for SiC$\{0001\}$ surfaces
and graphitic overlayers at SiC$\{0001\}$ surfaces.
In particular, we focus on surface and surface/overlayer stability,
and on binding and band-structure modifications (due to the binding)
of the overlayers.

For surfaces, we study the relative stability 
as function of the detailed stacking,
as function of the chemical composition and
to some extent (see supplementary material),
as function of the type of reconstruction.
We find that the surface-energy differences due 
to different detailed stacking of the outermost surfaces
are below the accuracy of our calculations and 
not of any significance.
For ideally truncated surfaces (no excess or deficiency of Si or C)
we find that Si-terminated surfaces are generally more favorable
than C-terminated surfaces.

For SiC/overlayer systems we find two different types
of overlayers. 
At SiC$(0001)$, we predict an overlayer that is chemically 
bonded to the substrate.
Because of the chemical bonding this overlayer is expected 
to significantly differ in its electronic nature from 
single-layer graphene.
At SiC$(000\bar1)$, we predict a vdW-bonded overlayer.

In line with Ref.~\onlinecite{Jacobsen},
our band-structure calculations for the vdW-bonded graphene
show that vdW interactions with a substrate
can have a doping effect (here: p doping).
As a novel feature, we also identify a kink in the electron dispersion
at the Fermi level
and calculate an effective mass of 
$m_{\text{eff}}\sim1.1m_{\text{e}}$
at the minimum of the conduction band
at this kink.

\section*{Acknowledgments}
We thank T. S. Rahman and  S. Hong  for valuable discussions.
Support by the Swedish National Graduate School in Materials Science (NFSM),
the Swedish Research Council (VR),
the Swedish Governmental Agency for Innovation Systems (VINNOVA),
and the Swedish National Infrastructure for Computing (SNIC)
is gratefully acknowledged.

\clearpage

\section*{Relative stability of $6H$-SiC$\{0001\}$ surface terminations
and formation of graphene overlayers by Si evaporation: Supplementary material}

\begin{figure}
\begin{tabular}{cc}
Si-terminated $(0001)$ surfaces\\
\includegraphics[width=8cm]{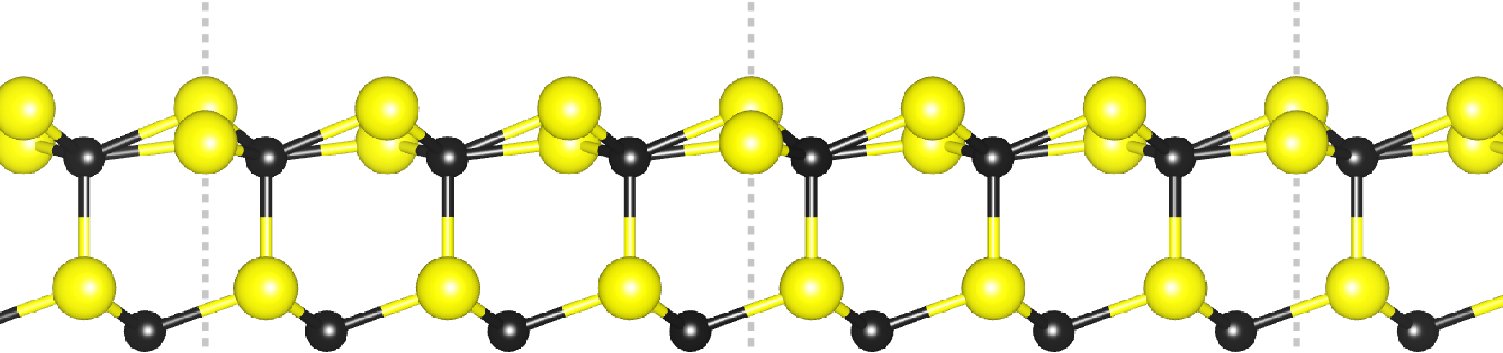}\\[0.2cm]
\includegraphics[width=8cm]{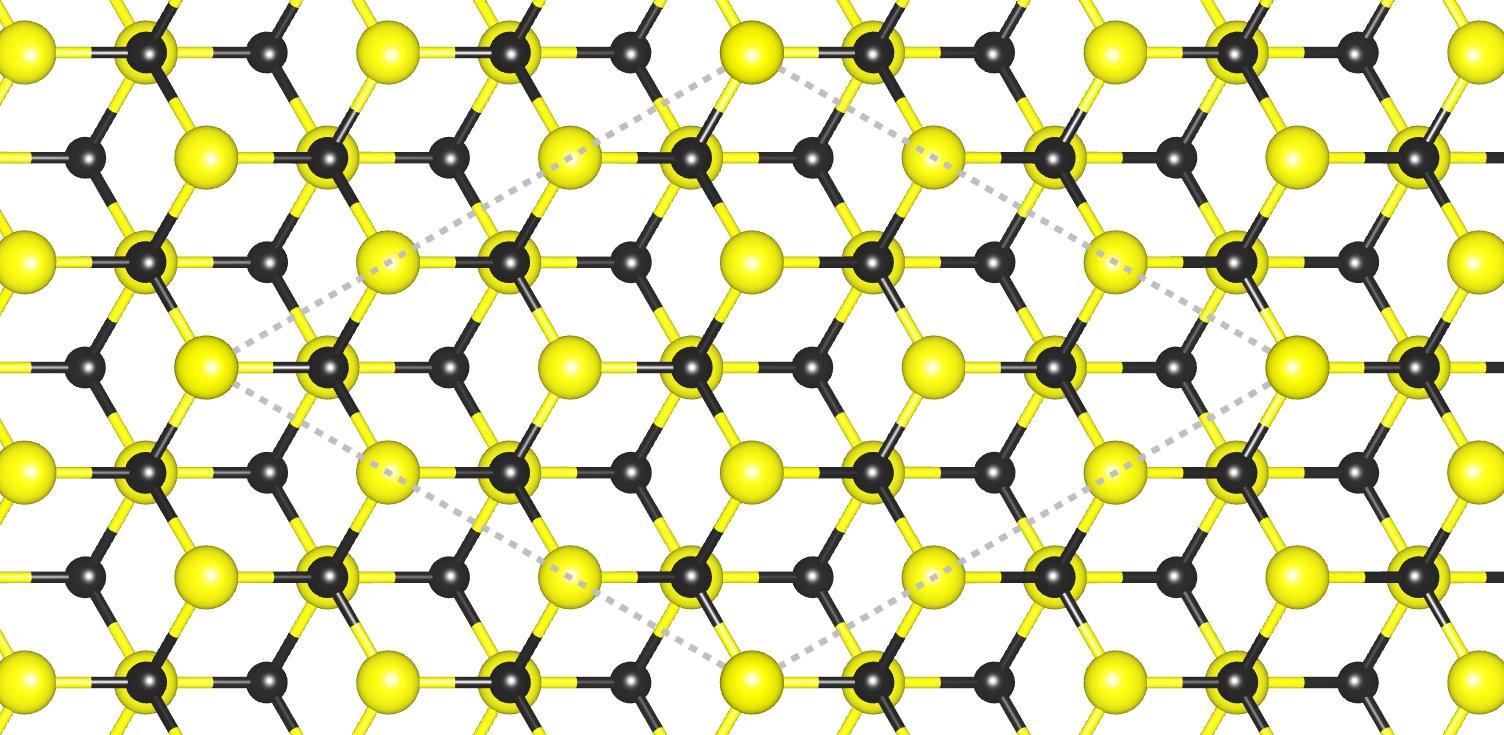}\\
[0.4cm]
C-terminated $(0001)$ surfaces\\
\includegraphics[width=8cm]{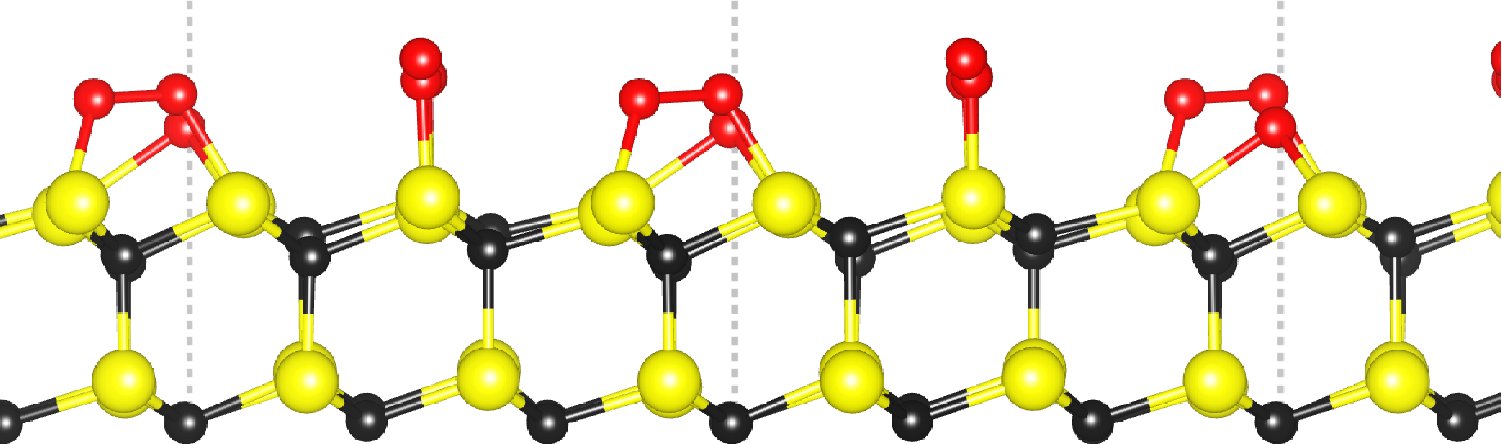}\\[0.2cm]
\includegraphics[width=8cm]{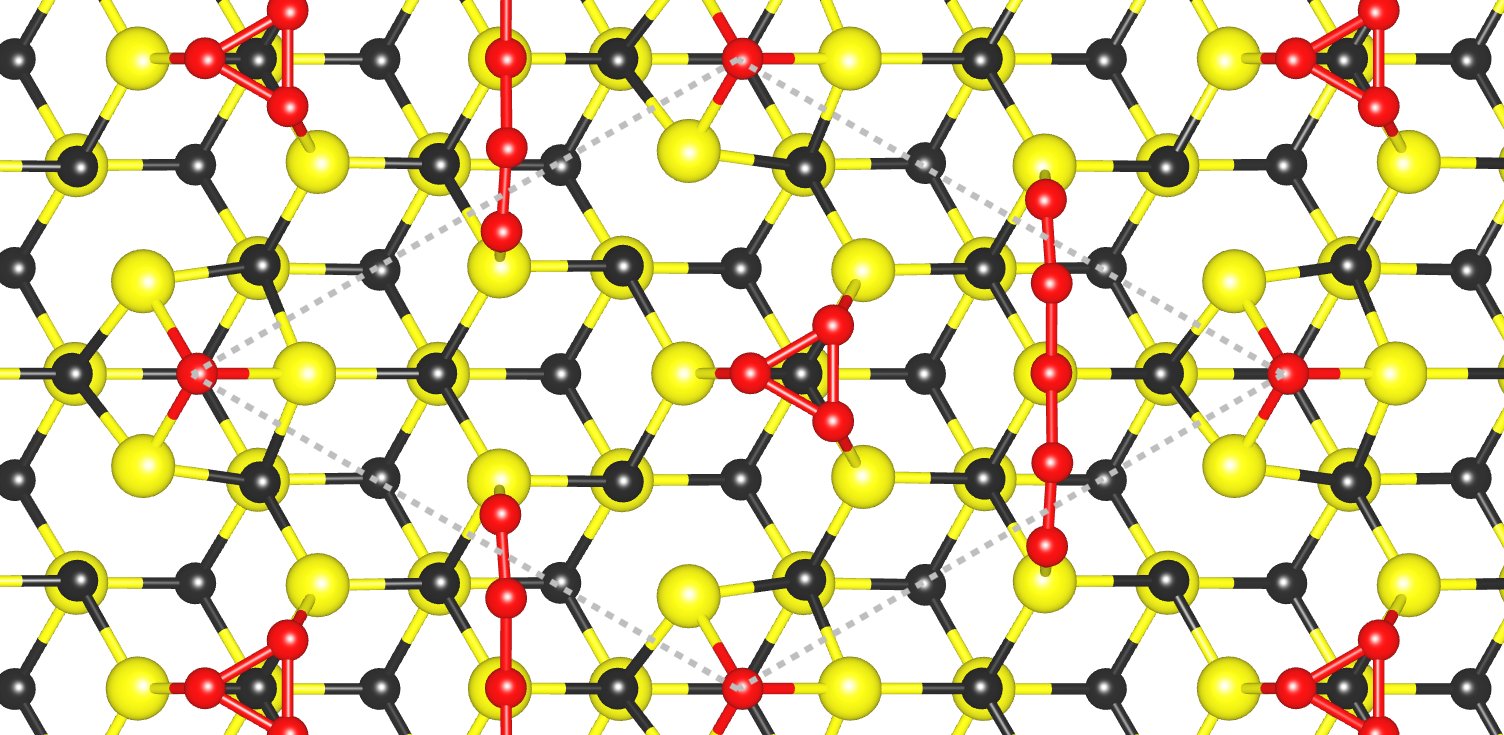}\\
[0.4cm]
Si-terminated $(000\bar1)$ surfaces\\
\includegraphics[width=8cm]{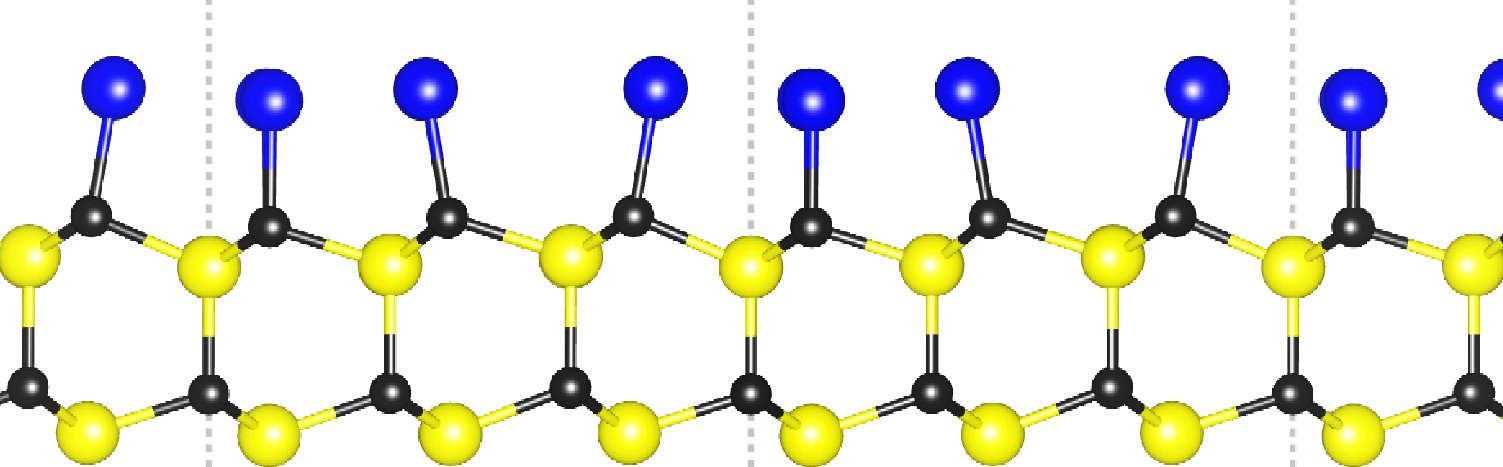}\\[0.2cm]
\includegraphics[width=8cm]{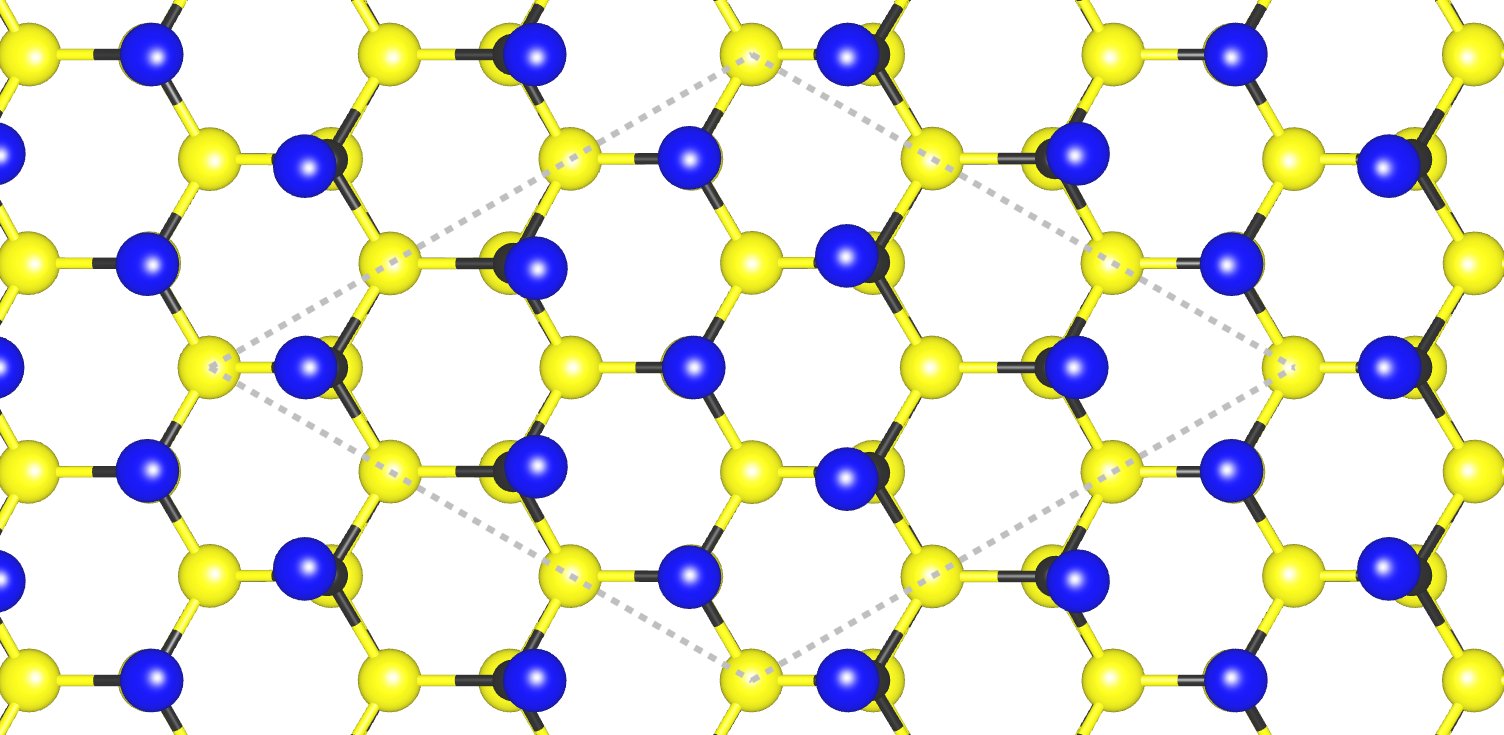}
\end{tabular}
\caption{
\label{fig:reconstructions}
Top and side views of $(3\times3)$ surface reconstructions
identified in this study.
Different colors are used to distinguish surface atoms that
undergo strong rearrangements from those in subsurface layers.
}
\end{figure}

In the main article entitled 'Relative stability of $6H$-SiC$\{0001\}$ surface terminations
and formation of graphene overlayers by Si evaporation',
we investigate, among others, the relative stability of $6H$-SiC$\{0001\}$ surfaces
as a function of their surface termination.
Focusing on $(1\times1)$ surface unit-cells,
we find that, independent of the orientation, Si termination
is preferred over C termination.
This result seems to be partially in conflict with intuition
based on counting the number of dangling bonds;
at SiC$(000\bar1)$, a C-terminated surface would be expected.

Here we present calculations where we 
consider the larger $(\sqrt3\times\sqrt3)$ and  $(3\times3)$ surface unit-cells.
These cells, in principle, allow for various surface reconstructions,
and the absence of such reconstructions in $(1\times1)$ surface unit-cells
could be the origin of the counterintuitive prediction
of a preferred Si termination at  SiC$(000\bar1)$.

For all calculations we use a common $400$~eV planewave cutoff.
For $(\sqrt3\times\sqrt3)$ surface cells, a $(3\times3\times1)$ 
k-point sampling is used;
for $(3\times3)$ surface cells, a $(2\times2\times1)$ k-point sampling
is used.
The surfaces are modeled as slabs consisting of at least six SiC bilayers.
All slabs are ideally truncated,
that is, they possess either a full-coverage Si-terminated surface
or a full-coverage C-terminated surface.
The surface models are relaxed until the forces on atoms no longer
exceeds $0.01$~eV/\AA.

Figure~\ref{fig:reconstructions} details different 
$(3\times3)$ surface reconstructions obtained from our calculations.
In fact, the direct relaxation of truncated surface slabs only produces
flat surfaces without true reconstructions,
the  Si-terminated $(3\times3)$ $(0001)$ surface being an exception.
All other reconstructions are obtained 
by pulling one atom out  of the flat surface-layer
by $0.5$~\AA\ and then restarting the relaxations.

The most pronounced reconstruction, giving rise to
triangular features, is identified 
at the C-terminated $(3\times3)$ $(0001)$ surface, see set of mid-panels in Fig.~\ref{fig:reconstructions}.
At Si-terminated $(3\times3)$ $(0001)$, see set of top panels in Fig.~\ref{fig:reconstructions}, 
and Si-terminated $(3\times3)$ $(000\bar1)$, see set of bottom panels in Fig.~\ref{fig:reconstructions},
only smaller departures from the flat surface are found.
The C-terminated $(3\times3)$SiC$(000\bar1)$,
remains unreconstructed even after triggering reconstructions
by pulling atoms out of the surface.

For $(\sqrt3\times\sqrt3)$ surface unit-cells,
all but the Si-terminated $(0001)$ configuration 
remain flat.
The reconstruction of the  Si-terminated  
$(\sqrt3\times\sqrt3)$ $(0001)$ surface
is similar to the reconstruction of the 
Si-terminated $(3\times3)$ $(0001)$ surface
and therefore not shown.

The reconstructed configurations generally possess lower energies 
than their unreconstructed, flat counterparts.
These energies are  used in our comparison of surface
energies in Table III in the paper.

We are aware of the likely fact that the here-presented reconstructions
only represent a small sample of the rich SiC surface phase-diagram.
The $(\sqrt3\times\sqrt3)$ and  $(3\times3)$ surface unit-cells
allow for a large freedom in the variation of the coverage of surface
Si or C.
Closely related is the problem of identifying
the lowest-energy structure for each coverage.
A full reconstruction search would require 
more advanced strategies and methods
than used here and is outside the scope 
of the present work.

Finally, we notice that identification of the full spectrum of surface reconstructions
is in fact only of secondary relevance for the actual purpose
of the main paper:
the study of band-structure modifications
due to van der Waals binding of graphitic overlayers at SiC$\{0001\}$.
These overlayers arise from evaporation of Si atoms from SiC$\{0001\}$.
As discussed in the main paper, 
the detailed structure and composition of the surface prior
to evaporation may be only of minor relevance.

\end{document}